\documentclass[preprint,aps,prd,showpacs,nofootinbib]{revtex4}
\usepackage{mathrsfs}
\usepackage{amsmath}
\usepackage{graphicx}
\usepackage{subfigure}

\newcommand{\bea}{\begin{eqnarray}}
\newcommand{\eea}{\end{eqnarray}}
\newcommand{\beq}{\begin{equation}}
\newcommand{\eeq}{\end{equation}}

\def\x{{\vec x}}
\def\/{\over}
\makeatletter

\newcommand{\Rmnum}[1]{\expandafter\@slowromancap\romannumeral #1@}
\makeatother

\begin{document}

\title{Response of a uniformly accelerated detector to massless Rarita-Schwinger fields in vacuum}
\author{  Qinglin Li$^{1}$, Hongwei Yu$^{1,2,}$\footnote{Corresponding author}$^{,}$\footnote{email: hwyu@hunnu.edu.cn} and Wenting Zhou$^{2}$ }
\affiliation{$^1$ Department of Physics and Key Laboratory of Low
Dimensional Quantum Structures and Quantum
Control of Ministry of Education,\\
Hunan Normal University, Changsha, Hunan 410081, China \\
$^2$ Center for Nonlinear Science and Department of Physics, Ningbo
University,  Ningbo, Zhejiang 315211, China}

\begin{abstract}
We study the response of a uniformly accelerated detector modeled by a two-level atom nonlinearly coupled to
vacuum massless Rarita-Schwinger fields. We first generalize the formalism developed by Dalibard, Dupont-Roc, and Cohen-Tannoudji in the
linear coupling case, and we then calculate the mean rate of change
of the atomic energy of the accelerated atom. Our result shows that a uniformly accelerated atom in its ground state interacting with vacuum
Rarita-Schwinger field fluctuations would spontaneously transition to an excited state and
the unique feature in contrast to the case of the atom coupled to the scalar,
electromagnetic and Dirac fields is the appearance of terms in the excitation rate which are proportional to the sixth and eighth powers of acceleration.

\end{abstract}
%\pacs{04.62.+v, 42.50.Lc, 03.70.+k} 
\maketitle

\baselineskip=16pt

\section{Introduction}
%%%%%%%%%%%%%%%%%%%%%%%%%

Unruh discovered in1976, by examining the response of a so-called Unruh-DeWitt particle detector to massless scalar fields, that for a uniformly accelerated observer, the Minkowski vacuum is seen to be equivalent to a
thermal bath of Rindler particles at a temperature $T_U = a/2\pi$~\cite{Unruh}, where $a$ is the observer's proper acceleration.  Since then, the Unruh effect has attracted a great deal of attention both for the crucial role it has played in our
understanding  that the particle content of a quantum
field theory is observer dependent and for its close relationship to the Hawking radiation from black holes (see Refs.~\cite{Takggi,Pad,Matsas08} for reviews for extensive works on the Unruh effect and its applications).  Recently, the Unruh effect has been studied from the perspective of a two-level atom interacting with vacuum quantum fields, such as scalar~\cite{Audretsch94,Audretsch951,Audretsch952}, electromagnetic~\cite{Passante98,Zhu06,Yu06,Rizzuto07,Rizzuto091,Zhu10} and Dirac fields~\cite{ZhouYu12}, in  the formalism developed by Dalibard, Dupont-Roc, and Cohen-Tannoudji (DDC)~\cite{Dalibard82,Dalibard84}, where the atom's excitation is distinctively attributed to the contributions that result from fluctuations in the vacuum,
and those that are due to the disturbance of the quantum field caused by the atom-field coupling. These studies show that if the atom is accelerated, then the delicate balance between vacuum fluctuations and radiation reaction that ensures the stability of the ground-state inertial atoms in vacuum is altered, thereby making transitions to excited states for the ground-state atoms  possible even in vacuum. This result is not only consistent with the Unruh effect, but also provides a physically appealing interpretation of it,  since it gives a transparent illustration for why an accelerated detector clicks.

In this paper, we plan to go a step further to make the
spectrum of research more complete, that is, we will extend the studies on the Unruh effect to particle detectors coupled to quantum fields of higher spin. In particular, we will examine the spontaneous excitation of an accelerated detector modeled by a  two-level atom in interaction with vacuum spin $3/2$ fields, i.e., Rarita-Swhinger fields~\cite{rarita}, using the DDC formalism that makes use
of a well-defined separation of detector excitations into vacuum fluctuations and radiation reaction.
Our interest in the issue also lies in the role Rarita-Schwinger fields play in supersymmetry.  As it
is well-known, supersymmetry is a relativistic symmetry
between bosons and fermions~\cite{17,18,19}. It unites in a single supermultiplet
particles with different intrinsic spins differing by units of one half
so that fermions become superpartners of bosons and vice versa. If it is correct, supersymmetry implies that every known
elementary particle must have a superpartner. So,   the graviton of spin $2$ supposed to mediate gravitation, one of the four fundamental interactions in nature, has a superpartner called  the gravitino which is described a spin-$3\/2$  field, i.e., the Rarita-Schwinger
field. We plan to calculate the spontaneous excitation rate of a uniformly accelerated  two-level atom in interaction with  vacuum Rarita-Schwinger fields using the DDC formalism.

The paper is organized as follows. In Sec. II, we generalize the DDC
formalism to Rarita-Schwinger fields where the coupling between an
atom and the field is nonlinear and then in Sec. III, we use the
generalized DDC formalism to calculate the spontaneous excitation
rate of uniformly accelerated atoms interacting with fluctuating
Rarita-Schwinger fields in vacuum. We conclude in Sec. IV. Natural
units in which $\hbar=c=1$ and  metric signature $(+,-,-,-)$ will be used
throughout the paper.

%%%%%%%%%%%%%%%%%%%%%%%%%%%%%
\section{the general formalism}

The system we shall consider consists of a two-level atom and a bath of vacuum fluctuating massless Rarita-Schwinger fields in four dimensional Minkowski spacetime. The two states of the atom, i.e., the ground and
excited states, are denoted by $|-\rangle$, $|+\rangle$,
respectively, with energies being $-{1\/2}\omega_0$ and
$+{1\/2}\omega_0$ . The coordinates of the atom associated with an inertial
reference are denoted by $x^{\mu}=(x^0,\x)=(t,x,y,z)$. The atom is
assumed to be on a stationary trajectory
$x(\tau)=(t(\tau),\x(\tau))$ where $\tau$ indicates the proper time.
The evolution of the atom in proper time $\tau$ is controlled  by the following Hamiltonian~\cite{Dicke},
 \beq
H_A(\tau)=\omega_0R_3(\tau)\label{atomic Hamiltonian}\;
 \eeq
where
$R_3(0)=\frac{1}{2}|+\rangle\langle+|-\frac{1}{2}|-\rangle\langle-|$.
The Lagrangian of the massless Rarita-Schwinger
field $\psi_\rho(x)$ which the atom is assumed to be coupled to is given by~\cite{22}
  \beq
 {\cal
L}=-\epsilon^{\lambda\rho\mu\nu}\bar{\psi}_\lambda\gamma_5\gamma_\mu\partial_\nu\psi_\rho\;,
\label{lar}
 \eeq
where
 \bea \gamma^0=\begin{pmatrix}
  I & 0 \\
  0 & -I \\
\end{pmatrix}\;,\quad
\gamma^i=\begin{pmatrix}
  0 & \mathbf{\sigma}_i \\
  -\mathbf{\sigma}_i & 0 \\
\end{pmatrix}
 \eea
 with $\sigma$ being the Pauli matrices,  $\gamma_5=i\gamma^0\gamma^1\gamma^2\gamma^3,
\epsilon^{0123}=1$, and
$\bar{\psi}_\lambda={\psi_\lambda}^\dagger\gamma^0$. The
$\gamma$-matrices
 satisfy the algebra: $\{\gamma^{\mu},\gamma^{\nu}\}=2g^{\mu\nu}$,
 where $\{,\}$ stands for the anticommutator.  The
 Lagrangian~(\ref{lar}) leads to the following  Euler-Lagrange equation of motion
  \beq
 i[\;g^{\lambda\rho}/\kern-0.50em\partial-\gamma^\lambda\partial^\rho
 -\gamma^\rho\partial^\lambda+\gamma^\lambda/\kern-0.50em\partial\gamma^\rho\;]\psi_\rho=0\;,\label{motion}
 \eeq
 where $/\kern-0.50em\partial=\gamma^\alpha\partial_\alpha$.
This equation is invariant under chiral rotations as well as
fermionic gauge transformations
$\psi_\rho\rightarrow\psi_\rho+\partial_\rho\phi$, where $\phi$ is
an arbitrary spinor field. To fix the gauge, we impose the following
conditions~\cite{das}
 \beq
 \gamma^i\psi_i=0\;, \quad\quad
 \psi_0=0\;, \quad\quad
 \partial^i\psi_i=0\;.
  \label{gauge}
 \eeq
% which plays the same role as the Coulomb gauge condition for electromagnetic fields.
%A contraction of (\ref{motion}) with $\gamma_\lambda$ yields
 % $
  %\partial^\rho\psi_\rho-/\kern-0.50em\partial(\gamma^\rho\psi_\rho)=0\;,\label{con3}
  %$
%which becomes, after the gauge condition~(\ref{gauge}) is used
% At the same time,  Eq~(\ref{motion}) for $\lambda=0$
 %gives a primary constraint equation,
  %$
  %\partial^i\psi_i+(\gamma^i\partial_i)(\gamma^i\psi_i)=0\;.
  %$
%Applying the gauge condition~(\ref{gauge}) again, one
%finds
 Then the original Euler-Lagrange
equation becomes the Dirac-like equation of motion,
 %\beq
 $i/\kern-0.50em\partial\psi_\mu=0\;.$
 %\eeq
Consequently, the Rarita-Schwinger field has a general plane-wave expansion~\cite{das},
 \beq
 \psi_{\mu}(x)=\sum_{s=\pm{3\/2}}\int{d^3\vec{p}\/\sqrt{(2\pi)^{3}\;2p_0}}[c(\vec{p},s,t)\;u_\mu(\vec{p},s)e^{-ip\cdot
 x}+d^*(\vec{p},s,t)\;v_\mu(\vec{p},s)e^{ip\cdot
 x}]\;.\label{expansion}
 \eeq
Here, the mode function $u_\mu(\vec{p},s)$ is given by
 \beq
 u_\mu(\vec{p},\pm{3\/2})={\cal D}_3(\phi){\cal D}_2(\theta)u_\mu(p_0,\pm{3\/2})\;,
\eeq
 with
 \beq
 u_\mu(p_0,\pm{3\/2})=\epsilon_\mu(p_0,\pm)u(p_0,\pm)\;,
 \eeq
 where the polarization vectors, $\epsilon_\mu(p_0,\pm)$, and the Dirac spinors in the direction $p^\mu=(p_0, 0, 0, p_0)$ (
 $p_0>0$), $u(p_0,\pm)$, are given respectively by
\beq
 \epsilon_\mu(p_0,\pm)=\mp\sqrt{{1\/2}}(0,1,\pm i,0)\;,
 \eeq
 and
 \beq
 u(p_0,\pm)=\sqrt{p_0}
 \begin{pmatrix}
  \varphi_\pm \\
  \pm\varphi_\pm\\
 \end{pmatrix}\;\text{with}\;
 \varphi_+=
 \begin{pmatrix}
 1 \\
 0 \\
 \end{pmatrix}\;,\;
 \varphi_-=
 \begin{pmatrix}
 0 \\
 1 \\
 \end{pmatrix}\;.
 \eeq
  ${\cal D}_3(\phi){\cal D}_2(\theta)$ is the
rotation of $u_\mu(\vec{p},\pm{3\/2})$ from the
 direction $p^\mu=(p_0, 0, 0, p_0)$ to a generic direction $p^\mu=(p_0,
p_0\sin\theta\cos\phi, p_0\sin\theta\sin\phi, p_0\cos\theta)=(p_0,
\vec{p})$. The mode function
 $v_\mu(\vec{p},s)$ can be obtained by the relation
 $v_\mu(\vec{p},s)=C\bar{u}^T_\mu(\vec{p},s)$, where
 $C=i\gamma^2\gamma^0$ and $\bar{u}_\mu(\vec{p},s)=u^\dag_\mu(\vec{p},s)\gamma^0$. These mode functions are normalized such that
  \beq
 \bar{u}^\mu(\vec{p},s')\gamma^\nu u_\mu(\vec{p},s)=
 \bar{v}^\mu(\vec{p},s')\gamma^\nu v_\mu(\vec{p},s)=
 -\delta_{ss'}2p^\nu \label{14}
  \eeq
  and
\beq
 \bar{u}^\mu(\vec{p},s')\gamma^0 v_\mu(\vec{p},s)=0\;.
  \eeq
Here the anti-commutation relations of $c(\vec{p},s)$ and
$d^*(\vec{p},s)$ as fermion creators and annihilators are
 \bea
  \{c(\vec{p},s),c^*(\vec{p}\;',s')\}=\delta^3(\vec{p}-\vec{p}\;')\delta_{ss'}
 =\{d(\vec{p},s),d^*(\vec{p}\;',s')\}\;,\nonumber\\
 \{c(\vec{p},s),c(\vec{p}\;',s')\}=\{d(\vec{p},s),d(\vec{p}\;',s')\}
 =\{c(\vec{p},s),d(\vec{p}\;',s')\}=0\;. \label{antirelation}
 \eea
The vacuum of Rarita-Schwinger fields is defined by the annihilation
operators as
 \beq
c(\vec{p},s)|0\rangle=d(\vec{p},s)|0\rangle=0\;.\label{vacuum}
 \eeq
The evolution of free
Rarita-Schwinger fields in proper time $\tau$  is generated by the free Hamiltonian
given by
 \beq
H_F(\tau)=\sum_s\int
   d^3\vec{p}\;\omega_{\vec{p}}\;[c^*(\vec{p},s)c(\vec{p},s)+d^*(\vec{p},s)d(\vec{p},s)]{dt\/d\tau}\;.
 \eeq
The atom interacts with vacuum Rarita-Schwinger fields and we assume the interaction Hamiltonian to be
 \bea
H_I(\tau)=\mu
R_2(\tau)\partial_\alpha{\bar{\psi}^\mu}(x(\tau))\partial^\alpha\psi_\mu(x(\tau))\;,\label{hi}
\eea
where $\bar{\psi}^\mu(x(\tau))={\psi^\mu}^\dagger(x(\tau))\gamma^0$, $\mu$
is the coupling constant that is assumed to be small and
$R_2(0)=\frac{1}{2}i[R_-(0)-R_+(0)]$ with
$R_+(0)=|+\rangle\langle-|$ and $R_-(0)=|-\rangle\langle+|$ being
the atomic raising and lowering operators respectively. These
operators obey $[R_3,R_\pm]=\pm
R_\pm$, and $[R_+,R_-]=2R_3$ .
Similar to the case of Dirac fields\cite{ZhouYu12}, the interaction Hamiltonian here is also quadratic in the field
operator. This type of nonlinear interaction  differs remarkably from  the linear atom-field couplings in the scalar and electromagnetic field cases~\cite{Audretsch94,Audretsch951,Audretsch952,Passante98,Zhu06,Yu06,Rizzuto07,Rizzuto091,Zhu10} in the sense that it makes  atomic transitions
 via both absorption and emission of Rarita-Schwinger
particle-antiparticle pairs  and inelastic scattering of a particle
or antiparticle possible at the lowest order of perturbation, whereas in the linear coupling case,
the quantum is singly absorbed or emitted and inelastic scattering
occurs only at higher orders.

Now, with the total Hamiltonian of the system (atom+field) which is given by: $
H(\tau)=H_A(\tau)+H_F(\tau)+H_I(\tau)$, we can derive the Heisenberg
equations of motion for the dynamical variables of the atom and the
field, and their
solutions can be separated into two
parts: the free part (denoted by index \textquotedblleft
f\textquotedblright) that is present even if no coupling between the atom and the field is assumed to exist
 and the source part (denoted by index
\textquotedblleft s\textquotedblright) characterized by the coupling constant $\mu$, which is generated
by the interaction. They are associated with vacuum fluctuations and
radiation reaction, respectively. If we assume that the state of the system is $|0,b\rangle$, where $0$
represents the vacuum state of the field and $b$ the state of the atom, and choose a symmetric ordering
between atom and field variables~\cite{Dalibard82,Dalibard84}, then following the same procedure as that in Ref.~\cite{ZhouYu12},
 we find, to the order $\mu^2$, the mean rate of change
of the atomic energy
 \beq
\biggl<{dH_A(\tau)\/d\tau}\biggl>=\biggl<{dH_A(\tau)\/d\tau}\biggr>_{vf}+\biggl<{dH_A(\tau)\/d\tau}\bigg>_{cross}
\label{decompose}\;,
 \eeq
where
 \bea
\biggl\langle{dH_A(\tau)\/d\tau}\biggr\rangle_{vf}
&=&{1\/2}i\mu\omega_0\langle\partial_\alpha\bar{\psi}^{\mu
f}(x(\tau))\partial^\alpha\psi^{\;\;f}_\mu(x(\tau))[R_2(\tau),R_{3}(\tau)]\nonumber
\\&&+[R_2(\tau),R_{3}(\tau)]\partial_\alpha\bar{\psi}^{\mu f}(x(\tau))\partial^\alpha\psi^{\;\;f}_\mu(x(\tau))\rangle\;,\label{vf1} \\
\biggl\langle{dH_A(\tau)\/d\tau}\biggr\rangle_{cross}\nonumber
&=&{1\/2}i\mu\omega_0\langle\partial_\alpha\bar{\psi}^{\mu
f}(x(\tau))\partial^\alpha\psi^{\;\;s}_\mu(x(\tau))[R_2(\tau),R_{3}(\tau)]\nonumber
\\&&+[R_2(\tau),R_{3}(\tau)]\partial_\alpha\bar{\psi}^{\mu f}(x(\tau))\partial^\alpha\psi^{\;\;s}_\mu(x(\tau))\nonumber\\
&&+\partial_\alpha\bar{\psi}^{\mu
s}(x(\tau))\partial^\alpha\psi^{\;\;f}_\mu(x(\tau))[R_2(\tau),R_{3}(\tau)]\nonumber
\\&&+[R_2(\tau),R_{3}(\tau)]\partial_\alpha\bar{\psi}^{\mu
s}(x(\tau))\partial^\alpha\psi^{\;\;f}_\mu(x(\tau))\rangle\;.\label{rr1}
 \eea
Notice that the cross-term  contains both
the free part and the source part and it is of the same order as the sole vacuum fluctuation term in our perturbative treatment correct to the order
$\mu^2$.  The cross-term dominates however over the sole radiation reaction term
which is of order $\mu^3$ and which can thus be neglected.  This term does not
 appear at the order $\mu^2$ in the cases of linear couplings~\cite{Audretsch94,Audretsch951,Audretsch952,Passante98,Zhu06,Yu06,Rizzuto07,Rizzuto091,Zhu10}.  The above equations can be further simplified to
 \bea
\biggl\langle{dH_A(\tau)\/d\tau}\biggr\rangle_{vf}& =&2i
\mu^2\int^{\tau}_{\tau_0}d\tau'\;C^F(x(\tau),x(\tau')){d\/d\tau}\chi^A(\tau,\tau')\;,
\label{general vf contribution}\\
\biggl\langle{dH_A(\tau)\/d\tau}\biggr\rangle_{cross}&
=&2i\mu^2\int^{\tau}_{\tau_0}d\tau'\;\chi^F(x(\tau),x(\tau')){d\/d\tau}C^A(\tau,\tau')\;.
\label{general rr contribution}
 \eea
Here $C^F(x(\tau),x(\tau'))$ and $\chi^F(x(\tau),x(\tau'))$ are the two statistical functions of the field defined respectively as
 \bea
C^F(x(\tau),x(\tau'))=C^F_+(x(\tau),x(\tau'))+C^F_-(x(\tau),x(\tau'))\;\label{c}\;,\nonumber
 \eea
 and
 \bea
\chi^F(x(\tau),x(\tau'))=\chi^F_+(x(\tau),x(\tau'))+\chi^F_-(x(\tau),x(\tau'))\;,\label{chi}
\nonumber \eea
with
 \bea
 C^F_+(x(\tau),x(\tau'))&=&\frac{1}{2}\langle0|\partial_\alpha\bar{\psi}^{\mu f}(x(\tau))\partial^\alpha\psi^{\;\;f}_\mu(x(\tau))
 \partial'_\beta\bar{\psi}^{\nu f}(x(\tau'))\partial'^\beta\psi^{\;\;f}_\nu(x(\tau'))|0\rangle\;,\\
 C^F_-(x(\tau),x(\tau'))&=&\frac{1}{2}\langle0|\partial'_\beta\bar{\psi}^{\nu f}(x(\tau'))\partial'^\beta\psi^{\;\;f}_\nu(x(\tau'))
 \partial_\alpha\bar{\psi}^{\mu f}(x(\tau))\partial^\alpha\psi^{\;\;f}_\mu(x(\tau))|0\rangle\;\;, \nonumber
 \eea
 and
 \bea
\chi^F_+(x(\tau),x(\tau'))&=&-\frac{1}{2}\langle0|\partial_\alpha\bar{\psi}^{\mu
f}(x(\tau))
[\partial'_\beta\bar{\psi}^{\nu f}(x(\tau'))\partial'^\beta\psi^{\;\;f}_\nu(x(\tau')),\partial^\alpha\psi^{\;\;f}_\mu(x(\tau))]|0\rangle\;,\\
\chi^F_-(x(\tau),x(\tau'))&=&-\frac{1}{2}\langle0|
[\partial'_\beta\bar{\psi}^{\nu
f}(x(\tau'))\partial'^\beta\psi^{\;\;f}_\nu(x(\tau')),
\partial_\alpha\bar{\psi}^{\mu f}(x(\tau))]\partial^\alpha\psi^{\;\;f}_\mu(x(\tau))|0\rangle\;,
 \nonumber
 \eea
 while $C^A(\tau,\tau')$ and $\chi^A(\tau,\tau')$ are the two susceptibility functions of the atom
 \bea
C^A(\tau,\tau')&=&\frac{1}{2}\langle
b|\{R_2^f(\tau),R_2^f(\tau')\}|b\rangle\;,\\
\chi^A(\tau,\tau')&=&\frac{1}{2}\langle
b|\;[R_2^f(\tau),R_2^f(\tau')\;]|b\rangle\;,
 \eea
 which can be explicitly written as
 \bea
C^A(\tau,\tau')&=&\frac{1}{2}\sum_{d}|\langle b|R_2(0)|d\rangle|^2\,
   (e^{i\omega_{bd}(\tau-\tau')}+e^{-i\omega_{bd}(\tau-\tau')})\;,\label{ca}\\
\chi^A(\tau,\tau')&=&\frac{1}{2}\sum_{d}|\langle
b|R_2(0)|d\rangle|^2\,
   (e^{i\omega_{bd}(\tau-\tau')}-e^{-i\omega_{bd}(\tau-\tau')})\;.\label{chia}
 \eea
The summation in the above two equations runs over the complete
set of the states of the atom.

\section{the spontaneous excitation in vacuum with Rarita-Schwinger field fluctuations}
Let us  assume that the two-level atom be uniformly accelerated such that its  trajectory is described by
 \beq
t(\tau)=\frac{1}{a}\sinh(a\tau)\;,\quad\;x(\tau)
=\frac{1}{a}\cosh(a\tau)\;,\quad\;y(\tau)=z(\tau)=0\label{traj}\;,
 \eeq
where $a$ is the proper acceleration.  Now we begin to calculate the  mean rate of  change of
the atomic energy of the uniformly accelerated atom using the formalism given in the preceding section.
To do this,  we need to compute the statistical functions
of Rarita-Schwinger fields. To simplify our computation, we
define a matrix related to the two-point function of Rarita-Schwinger fields as follows:
 \beq
S^+_{\mu\nu}(x(\tau),x(\tau'))=\langle0|\psi_\mu(x(\tau))\bar{\psi}_\nu(x(\tau'))|0\rangle\;.
\label{sp} \eeq
 Similarly, we can also define another one
$S^-_{\mu\nu}(x(\tau),x(\tau'))$ by:
 \beq
[S^-_{\mu\nu}(x(\tau),x(\tau'))]_{ab}=\langle0|(\bar{\psi}_\nu(x(\tau')))_b(\psi_\mu(x(\tau)))_a|0\rangle\;.\label{sg}
\eeq These two matrices are positive and negative frequency Wightman
functions of Rarita-Schwinger fields, respectively. Substituting
Eq.~(\ref{expansion}) into Eq.~(\ref{sp}), and utilizing
Eqs.~(\ref{14}) -~({\ref{vacuum}), we obtain
 \bea
 S^+_{\mu\nu}(x(\tau),x(\tau'))&=&\sum_{s=\pm{3\/2}}\int{d^3\vec{p}\/{(2\pi)^3\;2p_0}}u_\mu(\vec{p},s)\bar{u}_\nu(\vec{p},s)e^{-ip(x-x')}\nonumber\\
%&=&\int{d^3\vec{p}\/{(2\pi)^3\;2p_0}}[(\epsilon_\mu(\vec{p},+)u(\vec{p},+))(\epsilon^*_\mu(\vec{p},+)\bar{u}(\vec{p},+))\nonumber\\
%&&\;\;\;\;\;\;\;\;\;~~~~~~~~+(\epsilon_\mu(\vec{p},-)u(\vec{p},-))(\epsilon^*_\mu(\vec{p},-)\bar{u}(\vec{p},-))]e^{-ip(x-x')}\nonumber\\
%&=&\int{d^3\vec{p}\/{(2\pi)^3\;2p_0}}[-{i\/2}\;\bar{\delta}_{\nu\alpha}(\gamma_\alpha/\kern-0.50em\partial\gamma_\beta)\bar{\delta}_{\beta\mu}]e^{-ip(x-x')}\nonumber\\
&=&-{i\/2}\;\bar{\delta}_{\nu\alpha}(\gamma_\alpha/\kern-0.50em\partial\gamma_\beta)\bar{\delta}_{\beta\mu}G^+(x(\tau),x(\tau')),\label{s+}
 \eea
 where
\bea
\bar{\delta}_{\mu\nu}&=&\epsilon_\mu(\vec{p},+)\epsilon^*_\nu(\vec{p},+)+\epsilon_\mu(\vec{p},-)\epsilon^*_\nu(\vec{p},-)
=\delta_{\mu\nu}-(\partial_\mu\bar{\partial}_\nu+\partial_\nu\bar{\partial}_\mu)(\partial\bar{\partial})^{-1}
 \eea
with $\bar{\partial}_\mu=(-\partial_0,\partial_i)$,
$\partial\bar{\partial}=\partial^\alpha\bar{\partial}_\alpha$. In
the Coulomb-like gauge~(\ref{gauge}), there are no unphysical
modes  in the function above. Similarly, one can also show that
 \beq
 S^-_{\mu\nu}(x(\tau),x(\tau'))={i\/2}\;\bar{\delta}_{\nu\alpha}(\gamma_\alpha/\kern-0.50em\partial\gamma_\beta)\bar{\delta}_{\beta\mu}G^-(x(\tau),x(\tau')).\label{s-}
 \eeq
Here $G^+(x(\tau),x(\tau'))$ and $G^-(x(\tau),x(\tau'))$ are
positive and negative frequency Wightman functions of massless scalar
fields in four dimensional Minkowski spacetime, which are given respectively by
 \bea
G^+(x(\tau),x(\tau'))=-{1\/4{\pi}^2}{1\/(\triangle\tau-i\varepsilon)^2-|\vec{x}-\vec{x}\;'|^2}\;,\nonumber\\
G^-(x(\tau),x(\tau'))=-{1\/4{\pi}^2}{1\/(\triangle\tau+i\varepsilon)^2-|\vec{x}\;'-\vec{x}|^2}\;,\label{g}
 \eea
 where
$\triangle\tau=\tau-\tau'$ and $\varepsilon$ is a real infinitesimal
quantity. Exchanging  $x(\tau)$ and $x(\tau')$ in
$S^+_{\mu\nu}(x(\tau),x(\tau'))$ leads to
 \beq
S^+_{\mu\nu}(x(\tau'),x(\tau))={i\/2}\;\bar{\delta}_{\nu\alpha}(\gamma_\alpha/\kern-0.50em\partial\gamma_\beta)\bar{\delta}_{\beta\mu}G^+(x(\tau'),x(\tau))\;.
 \eeq
 Thus, noticing that $G^-(x(\tau),x(\tau'))=G^+(x(\tau'),x(\tau))$, we arrive at the relation:
 \beq
S^-_{\mu\nu}(x(\tau),x(\tau'))=S^+_{\mu\nu}(x(\tau'),x(\tau))\;.
\label{relation}
 \eeq

With the Wightman functions of massless Rarita-Schwinger
fields derived, now we can calculate the two statistical functions needed in our calculation of the mean
rate of change of the atomic energy. Using Eqs.~(\ref{sp})
and~(\ref{sg}), we see that $C^F_+(x(\tau),x(\tau'))$ and
$C^F_-(x(\tau),x(\tau'))$ can be written respectively as:
 \bea
C^F_+(x(\tau),x(\tau'))
=\text{Tr}[\partial^\alpha{\partial'}_\beta{S^{\;\nu}_\mu}^+(x(\tau),x(\tau'))\;
\partial_\alpha\partial'^\beta{S^{\;\mu}_{\nu}}^-(x(\tau'),x(\tau))]\;,\label{s0}
 \eea
\bea C^F_-(x(\tau),x(\tau'))
=\text{Tr}[\partial_\alpha\partial'^\beta{S^{\;\mu}_{\nu}}^-(x(\tau'),x(\tau))\;
\partial^\alpha{\partial'}_\beta{S^{\;\nu}_\mu}^+(x(\tau),x(\tau'))]\;,\label{s1}
 \eea
 and $\chi^F_+(x(\tau),x(\tau'))$ and $\chi^F_-(x(\tau),x(\tau'))$ respectively
 as:
 \bea
 \chi^F_+(x(\tau),x(\tau')) =-\text{Tr}[\partial_\alpha\partial'^\beta{S^{\;\mu}_\nu}^-(x(\tau'),x(\tau))\;
\partial'_\beta\partial^\alpha{S^{\;\nu}_\mu}^-(x(\tau),x(\tau'))]\;\;\nonumber\\
\;\;\;\;\;\;-\text{Tr}[\partial_\alpha\partial'^\beta{S^{\;\mu}_\nu}^-(x(\tau'),x(\tau))\;
\partial^\alpha\partial'_\beta{S^{\;\nu}_\mu}^+(x(\tau),x(\tau'))]\;,\label{s2}
 \eea
 \bea
\chi^F_-(x(\tau),x(\tau'))
=\text{Tr}[\partial'_\beta\partial^\alpha{{S^{\;\nu}_\mu}}^-(x(\tau),x(\tau'))\;
\partial'_\beta\partial^\alpha\;{S^{\;\mu}_{\nu}}^+(x(\tau'),x(\tau))]\;\;\;\;\;\nonumber\\
+\text{Tr}[\partial_\alpha\partial'^\beta{S^{\;\mu}_\nu}^-(x(\tau'),x(\tau))\;
\partial'_\beta\partial^\alpha{S^{\;\nu}_\mu}^-(x(\tau),x(\tau'))]\;,\label{s3}
 \eea
where $\text{Tr}[\cdots]$ represents the trace of a matrix.
Substituting Eqs.~(\ref{s+}) and~(\ref{s-}) into Eqs.~(\ref{s0})-~(\ref{s3}),
and using Eqs.~(\ref{g}), ~(\ref{relation}) and the trajectory~(\ref{traj}), we get the two statistical functions:
 \bea
C^F(x(\tau),x(\tau'))&=&-{183a^{10}\/32\pi^4}\biggl[{1\/\sinh^{10}({a\/2}\Delta\tau-i\varepsilon)}
+{1\/\sinh^{10}({a\/2}\Delta\tau+i\varepsilon)}\biggr]\;,\label{cf}\\
\chi^F(x(\tau),x(\tau'))&=&-{183a^{10}\/32\pi^4}\biggl[{1\/\sinh^{10}({a\/2}\Delta\tau-i\varepsilon)}
-{1\/\sinh^{10}({a\/2}\Delta\tau+i\varepsilon)}\biggr]\;,\label{chif}
 \eea

Inserting Eqs.~(\ref{cf}) and (\ref{chia}) into Eq.~(\ref{general vf
contribution}), extending the range of integration to infinity for
sufficiently long proper times, and
making use of the techniques of contour integration and residue theory,
we obtain the contribution of vacuum fluctuations to the mean
rate of change of the atomic energy:
 \bea
\biggl\langle{dH_A(\tau)\/d\tau}\biggr\rangle_{vf}
&=&-{61\mu^2\/1890\pi^3}\sum_{\omega_b>\omega_d}|\langle
b|R_2(0)|d\rangle|^2\omega_{bd}^{10}\nonumber\\
&&\;\;\;\;\times\biggl(1+30{a^2\/\omega_{bd}^2}+273{a^4\/\omega_{bd}^4}+820{a^6\/\omega_{bd}^6}+576{a^8\/\omega_{bd}^8}\biggr)
\biggl(1+{2\/{e^{2\pi\omega_{bd}/a}-1}}\biggr)\nonumber\\
&&\;+{61\mu^2\/1890\pi^3}\sum_{\omega_b<\omega_d}|\langle
b|R_2(0)|d\rangle|^2\omega_{bd}^{10}\nonumber\\
&&\;\;\;\;\times\biggl(1+30{a^2\/\omega_{bd}^2}+273{a^4\/\omega_{bd}^4}+820{a^6\/\omega_{bd}^6}+576{a^8\/\omega_{bd}^8}\biggr)
\biggl(1+{2\/{e^{2\pi|\omega_{bd}|/a}-1}}\biggr)\;. \nonumber\\
\label{final vf contribution}
 \eea
%(see Eq.~(56) in Ref.~\cite{10}, Eq.~(28) in Ref.~\cite{14} and Eq.~(64) in Ref.~\cite{16}),
This reveals that the contributions of vacuum
Rarita-Schwinger field fluctuations would increase the atomic energy when
the atom is initially in its ground state ( because of the term
$(\omega_b<\omega_d)$) and decrease it if otherwise (because of the term
$(\omega_b>\omega_d)$).  What distinguishes a uniformly accelerated atom in interaction with the Rarita-Schwinger field fluctuations from that in interaction with
massless scalar field, electromagnetic field and
Dirac field fluctuations is the appearance of extra terms which are proportional to $a^6$ and $a^8$.

Similarly, we can find, by plugging Eqs.~(\ref{chif}) and (\ref{ca})
into Eq.~(\ref{general rr contribution}), the contribution of the
cross term to the mean rate of change of the atomic energy:
  \bea
\biggl\langle{dH_A(\tau)\/d\tau}\biggr\rangle_{cross}
&=&-{61\mu^2\/1890\pi^3}\sum_{\omega_b>\omega_d}|\langle
b|R_2(0)|d\rangle|^2\omega_{bd}^{10}\nonumber\\
&&\;\;\;\;~~~~\times\biggl(1+30{a^2\/\omega_{bd}^2}+273{a^4\/\omega_{bd}^4}+820{a^6\/\omega_{bd}^6}+576{a^8\/\omega_{bd}^8}\biggr)\nonumber\\
&&\;-{61\mu^2\/1890\pi^3}\sum_{\omega_b<\omega_d}|\langle
b|R_2(0)|d\rangle|^2\omega_{bd}^{10}\nonumber\\
&&\;\;\;\;~~~~\times\biggl(1+30{a^2\/\omega_{bd}^2}+273{a^4\/\omega_{bd}^4}+820{a^6\/\omega_{bd}^6}+576{a^8\/\omega_{bd}^8}\biggr)\;. \nonumber\\
\label{final rr contribution}
 \eea
This tells us that the contribution of the cross term can only decrease the atomic energy no matter what
the initial state of the atom is.
Adding up Eqs.~(\ref{final vf contribution}) and (\ref{final rr
contribution}), we obtain the total mean rate of change of the
atomic energy
 \bea
 \biggl\langle{dH_A(\tau)\/d\tau}\biggr\rangle_{tot}
&=&-{61\mu^2\/945\pi^3}\sum_{\omega_b>\omega_d}|\langle
b|R_2(0)|d\rangle|^2\omega_{bd}^{10}\nonumber\\
&&\;\;\;\;\times\biggl(1+30{a^2\/\omega_{bd}^2}+273{a^4\/\omega_{bd}^4}+820{a^6\/\omega_{bd}^6}+576{a^8\/\omega_{bd}^8}\biggr)
\biggl(1+{1\/{e^{2\pi\omega_{bd}/a}-1}}\biggr)\nonumber\\
&&\;+{61\mu^2\/945\pi^3}\sum_{\omega_b<\omega_d}|\langle
b|R_2(0)|d\rangle|^2\omega_{bd}^{10}\nonumber\\
&&\;\;\;\;\times\biggl(1+30{a^2\/\omega_{bd}^2}+273{a^4\/\omega_{bd}^4}+820{a^6\/\omega_{bd}^6}+576{a^8\/\omega_{bd}^8}\biggr)
{1\/{e^{2\pi|\omega_{bd}|/a}-1}}\;. \nonumber\\\label{totol}
 \eea
As can be seen, from the above expression, the uniformly accelerated
atom in the ground state would spontaneously transition to an excited state in vacuum with
fluctuating Rarita-Schwinger fields. When $a\gg \omega_{bd}$, the terms proportional to $a^6$ and $a^8$ which distinguish the case  of the Rarita-Schwinger fields from the cases of scalar, electromagnetic and Dirac fields become preponderant over other terms. The appearance of power terms in $a$ in the mean rate of change of atomic energy for accelerated atoms in interaction
 with vacuum electromagnetic, Dirac, as well as Rarita-Schwinger fields we are discussing in
the present paper, which are absent in the case of a scalar field,  shows that the equivalence between
the acceleration and a thermal bath is lost in terms of the spontaneous excitation of atoms in these cases.
However, the deviation from
pure thermal behavior of the spontaneous excitation of the
uniformly accelerated atom by
no means implies the exact final thermal equilibrium is not
reached. In fact, with the
transition probabilities (excitation and emission) for the uniformly accelerated atom which
can be found from Eq.(\ref{totol}), one
can show, by the same argument as that in Ref~\cite{GP04},  that
exact thermal equilibrium will be established at the Unruh
temperature. Nevertheless, different behaviors of the transition
probabilities of the atoms do imply a clear difference  between a thermal bath and the acceleration
in terms of how atomic transitions occur and how the equilibrium is reached.

\section{Summary}

We have calculated, after generalizing the DDC formalism to the case of an atom coupled
nonlinearly to vacuum Rarita-Schwinger field fluctuations, the spontaneous
excitation of a uniformly accelerated atom in interaction with
fluctuating vacuum Rarita-Schwinger fields. Our result shows that
such a uniformly accelerated atom in its ground state would spontaneously transition to an excited state in
vacuum and the excitation rate contains, in addition to those associated with a pure thermal bath the Unruh temperature at
$T=a/2\pi$, terms proportional to the acceleration to the power of six and eight which are absent in the cases of the scalar, electromagnetic and Dirac fields. Although the behavior of a
uniformly accelerated atom deviates from that in a thermal bath in terms of the spontaneous excitation and emission due to the appearance of such power terms, the exact thermal equilibrium will still be established at the Unruh
temperature $a/2\pi$.  Nevertheless, different behaviors of the transition
probabilities of the atoms do imply a clear  distinguishability  between a thermal bath and the acceleration
in terms of how atomic transitions occur.

We must point out that the appearance in the excitation rate of terms of higher powers of acceleration compared to the case of Dirac fields~\cite{ZhouYu12}, i.e., terms proportional to the sixth and eighth powers of acceleration, is actually a result of the derivative coupling nature of the interaction. The same thing happens in the scalar field case, where powers of acceleration will appear in the excitation rate of a detector when a monopole coupling is replaced by a derivative coupling~\cite{Hinton83,Yu06,Matsas08,MML}. This is because of that the excitation rate depends on the derivatives of the Wightman function,  which increase the order of the pole in the $Sinh$ function. The higher the order of the pole, the higher the powers of acceleration in the excitation rate. It is also interesting to note that derivative coupling is not the sole reason for the increase in the order of the pole, in fact, any
perturbative expansion in a small parameter of the Wightman function, even for the monopole coupling case is sufficient(see, for example, Refs.~\cite{RT,KP}).

Finally, let us note that when one considers that the detector is switched on and off, then the choice of the form of the Wightman function Eq.~(34) in the present paper corresponds to a particular choice of the regularizaion function which requires smooth $C^\infty$ switching functions~\cite{Satz}. In this switching-on-and-off scenario, the switching functions considered by us here are actually Heavyside step functions which usually result in transient effects due to switching even for a stationary trajectory. The absence of such effects in the  present paper is however a result of the fact that we assume the detector to be switched on for an infinite time interval which makes the extra transient effects vanish.

\begin{acknowledgments}

This work was supported in part by the NNSFC under Grants No. 11075083 and No. 11375092; the National 973 Program of China under Grant No. 2010CB832803; the PCSIRT under Grant No. IRT0964; the Hunan Provincial Natural Science Foundation under Grant No. 11JJ7001 and the SRFDP under Grant No. 20124306110001
\end{acknowledgments}

\end{document}